\documentclass{revtex4}
\usepackage{amssymb}
\usepackage{amsfonts}
\usepackage{amsmath}

\input{tcilatex}

\begin{document}

\title{On the $n$-dimensional extension of Position-dependent mass
Lagrangians: nonlocal transformations, Euler--Lagrange invariance and exact
solvability}
\author{Omar Mustafa}
\email{omar.mustafa@emu.edu.tr}
\affiliation{Department of Physics, Eastern Mediterranean University, G. Magusa, north
Cyprus, Mersin 10 - Turkey,\\
Tel.: +90 392 6301378; fax: +90 3692 365 1604.}

\begin{abstract}
\textbf{Abstract:}\ The $n$-dimensional extension of the one dimensional
Position-dependent mass (PDM) Lagrangians under the nonlocal point
transformations by Mustafa \cite{38} is introduced. The invariance of the $n$%
-dimensional PDM Euler-Lagrange equations is examined using two
possible/different PDM Lagrangian settings. Under the nonlocal point
transformation of Mustafa \cite{38}, we have shown that the PDM
Euler-Lagrange invariance is only feasible for one particular
PDM-Lagrangians settings. Namely, when each velocity component is deformed
by some dimensionless scalar multiplier that renders the mass
position-dependent. Two illustrative examples are used as \emph{reference}
Lagrangians for different PDM settings, the nonlinear $n$-dimensional
PDM-oscillators and the nonlinear isotonic $n$-dimensional PDM-oscillators.
Exact solvability is also indulged in the process.

\textbf{PACS }numbers\textbf{: }05.45.-a, 03.50.Kk, 03.65.-w

\textbf{Keywords:} $n$-dimensional position-dependent mass Lagrangians,
nonlocal point transformation, Euler-Lagrange equations invariance.
\end{abstract}

\maketitle

\section{Introduction}

The mathematical challenge associated with the position-dependent mass (PDM)
von Roos Hamiltonian \cite{1}, and the feasible applicability of the PDM
settings in different fields of physics, has inspired a relatively intensive
recent research attention on the quantum mechanical (see the sample of
references \cite{2,3,4,5,6,7,8,9,10,11,12}), classical mechanical and
mathematical (see the sample of references \cite%
{12,13,14,15,16,17,18,19,20,21,22,23,24,25,26,27,28,29,30,31,32,33,34,35,36,37}%
) domains in general. The position-dependent mass, in principle, is either a
position-dependent deformation in the standard constant mass, or a
position-dependent deformation in the coordinates system, or even a
position-dependent deformation in the velocity components of the system.
Which, in turn, deforms the potential force field and may inspire nonlocal
space-time point transformations to facilitate exact solvability. Very
recently, Mustafa \cite{38} has introduced a nonlocal point transformation
for one-dimensional PDM Lagrangians and provided their mappings into a
constant \emph{"unit-mass"} Lagrangians in the generalized coordinates.
Therein, it has been shown that the applicability of such mappings not only
results in the linearization of some nonlinear oscillators but also extends
into the extraction of exact solutions of more complicated dynamical
systems. Hereby, the exactly solvable Lagrangians (labeled as \emph{%
"reference-Lagrangians"}) are mapped along with their exact solutions into
PDM-Lagrangians (labeled as \emph{"target-Lagrangians"}). It would be
natural and interesting, therefore, to extend and generalize Mustafa's
proposal \cite{38} to deal with Lagrangians in more than one-dimension.

In the current methodical proposal, we shall be interested in two types of $n
$-dimensional PDM systems. The first type of which has each velocity
component $\dot{x}_{_{j}}$ deformed by a dimensionless scalar multiplier $%
\sqrt{m_{_{i}}\left( x_{_{j}}\right) }$ to form the kinetic energy term 
\begin{equation}
T_{I}=\frac{1}{2}m_{\circ }\sum\limits_{j=1}^{n}m_{_{j}}\left(
x_{_{j}}\right) \dot{x}_{_{j}}^{2};\text{ }\dot{x}_{_{j}}=\frac{dx_{_{j}}}{dt%
};\text{ }\,\,\,j=1,2,\cdots ,n\in 
\mathbb{N}
.
\end{equation}%
The other type has all velocity component $\dot{x}_{_{j}}^{\prime }s$
deformed by a common dimensionless scalar multiplier $\sqrt{m\left( \vec{x}%
\right) }$ and forms a kinetic energy term%
\begin{equation}
T_{II}=\frac{1}{2}m_{\circ }\sum\limits_{j=1}^{n}m\left( \vec{x}\right) \dot{%
x}_{_{j}}^{2}.
\end{equation}%
Consequently, the corresponding potential energies shall be, respectively,
labeled as $V_{I}\left( \vec{x}\right) $ and $V_{II}\left( \vec{x}\right) $.
Likewise, we label the corresponding Lagrangians $L_{I}\left( 
\overrightarrow{x},\overrightarrow{\dot{x}};t\right) $ and $L_{II}\left( 
\overrightarrow{x},\overrightarrow{\dot{x}};t\right) .$

In section 2, we use the two types of the kinetic energy, (1) and (2), and
construct the corresponding Lagrangians $L_{I}\left( \overrightarrow{x},%
\overrightarrow{\dot{x}};t\right) $ and $L_{II}\left( \overrightarrow{x},%
\overrightarrow{\dot{x}};t\right) $ to workout their PDM Euler-Lagrange
equations of motion. We then use the usual textbook constant mass $m_{\circ }
$ Lagrangian $L\left( \overrightarrow{q},\overrightarrow{\tilde{q}};\tau
\right) $ in the generalized coordinates and buildup the corresponding
Euler-Lagrange equations in the generalized coordinates (to be identified as
EL-G). Moreover, within some extended form of the nonlocal point
transformation of Mustafa \cite{38} we try to search for feasible
Euler-Lagrange equations' invariance for the two systems. Under our extended
nonlocal point transformation\ we show that, whilst the PDM Euler-Lagrange
equations for $L_{I}\left( \overrightarrow{x},\overrightarrow{\dot{x}}%
;t\right) $ (to be denoted as "PDM EL-I") admit invariance with the EL-G,
the PDM Euler-Lagrange equations'  invariance for $L_{II}\left( 
\overrightarrow{x},\overrightarrow{\dot{x}};t\right) $ (to be denoted as
"PDM EL-II") is shown to be incomplete and still far beyond reach for $n\geq
2$. We, therefore, adopt the format of the PDM EL-I and proceed with our
proposal. In section 3, we use a set of nonlinear $n$-dimensional PDM
oscillators as illustrative examples. Amongst are the $n$-dimensional
Mathews-Lakshmanan type-I and type-II oscillators described, respectively,
by the PDM-Lagrangians (27) and (38) below. A power law type as well as an
exponential type $n$-dimensional PDM oscillators are also used in the same
section. Nonlinear isotonic $n$-dimensional PDM oscillators are reported in
section 4, where the PDM Smorodinsky-Winternitz type-I and type-II
oscillators are used as illustrative examples. We conclude in section 5.

\section{Nonlocal point transformations and multidimensional PDM
Euler-Lagrange invariance}

Let us consider two sets of multidimensional PDM-Lagrangians%
\begin{equation}
L_{I}\left( \overrightarrow{x},\overrightarrow{\dot{x}};t\right) =\frac{1}{2}%
m_{\circ }\sum\limits_{j=1}^{n}m_{_{j}}\left( x_{_{j}}\right) \,\dot{x}%
_{_{j}}^{2}-V_{I}\left( \vec{x}\right) ;\text{ }\dot{x}_{_{j}}=\frac{%
dx_{_{j}}}{dt};\text{ }\,\,\,j=1,2,\cdots ,n\in 
\mathbb{N}
,
\end{equation}%
and%
\begin{equation}
L_{II}\left( \overrightarrow{x},\overrightarrow{\dot{x}};t\right) =\frac{1}{2%
}m_{\circ }\sum\limits_{j=1}^{n}m\left( \vec{x}\right) \dot{x}%
_{_{j}}^{2}-V_{II}\left( \vec{x}\right) ,
\end{equation}%
where $m_{\circ }$ is the rest mass, $m_{_{j}}\left( x_{_{j}}\right) $ in $%
L_{I}\left( \overrightarrow{x},\overrightarrow{\dot{x}};t\right) $\ is a
dimensionless scalar multiplier that deforms each coordinate $x_{_{j}}$
and/or velocity component $\dot{x}_{_{j}}$ in a specific functional form,
and $m\left( \vec{x}\right) $ in $L_{II}\left( \overrightarrow{x},%
\overrightarrow{\dot{x}};t\right) $ represents a common dimensionless scalar
multiplier that deforms the coordinates $x_{_{j}}$'s and/or velocity
components $\dot{x}_{_{j}}$'s. Of course, a consequential position-dependent
deformation in the potential force fields $V_{I}\left( \vec{x}\right) $ and $%
V_{II}\left( \vec{x}\right) $ is unavoidable in this case. Under such
settings, both $L_{I}\left( \overrightarrow{x},\overrightarrow{\dot{x}}%
;t\right) $ and $L_{II}\left( \overrightarrow{x},\overrightarrow{\dot{x}}%
;t\right) $ may very well represent two different types of PDM Lagrangians,
so to speak.\ The Euler-Lagrange equations%
\begin{equation}
\frac{d}{dt}\left( \frac{\partial L}{\partial \dot{x}_{i}}\right) -\frac{%
\partial L}{\partial x_{i}}=0;\text{ }\,\,\,i=1,2,\cdots ,n\in 
\mathbb{N}
,
\end{equation}%
imply (with $m_{\circ }=1$ throughout for simplicity and economy of
notations) $n$ PDM Euler-Lagrange equations 
\begin{equation}
\ddot{x}_{_{i}}+\left( \frac{\dot{m}_{_{i}}\left( x_{i}\right) }{%
2m_{_{i}}\left( x_{i}\right) }\right) \,\dot{x}_{i}+\left( \frac{1}{%
m_{_{i}}\left( x_{i}\right) }\right) \,\partial _{x_{i}}V_{I}\left( \vec{x}%
\right) =0;\text{ }\ddot{x}_{_{j}}=\frac{d^{2}x_{_{j}}}{dt^{2}},
\end{equation}%
for $L_{I}\left( \overrightarrow{x},\overrightarrow{\dot{x}};t\right) $ (and
shall be referred to as "PDM EL-I", hereinafter), and%
\begin{equation}
\ddot{x}_{_{i}}+\left( \frac{\dot{m}\left( \vec{x}\right) }{m\left( \vec{x}%
\right) }\right) \dot{x}_{i}-\frac{1}{2}\left( \frac{\partial
_{x_{i}}m\left( \vec{x}\right) }{m\left( \vec{x}\right) }\right)
\sum\limits_{j=1}^{n}\dot{x}_{j}^{2}+\left( \frac{1}{m\left( \vec{x}\right) }%
\right) \,\partial _{x_{i}}V_{II}\left( \vec{x}\right) =0,
\end{equation}%
for\ $L_{II}\left( \overrightarrow{x},\overrightarrow{\dot{x}};t\right) $
(and shall be referred to as "PDM EL-II", hereinafter), where $\partial
_{x_{i}}=\partial /\partial x_{i}$.

On the other hand, consider a classical particle of a constant mass $%
m_{\circ }$ moving in a potential force field $V(\vec{q})$, where $\vec{q}%
=\left( q_{_{1}},q_{_{2}},\cdots ,q_{_{n}}\right) $ are some generalized
coordinates. The corresponding Lagrangian for such a system is given by%
\begin{equation}
L\left( \overrightarrow{q},\overrightarrow{\tilde{q}};\tau \right) =\frac{1}{%
2}m_{\circ }\sum\limits_{j=1}^{n}\tilde{q}_{_{j}}^{2}-V(\vec{q});\text{ \ }%
\tilde{q}_{_{j}}=\frac{dq_{_{j}}}{d\tau };\text{ }\,j=1,2,\cdots ,n,
\end{equation}%
where $\tau $ is a re-scaled time \cite{38}. The\ corresponding $n$
Euler-Lagrange equations%
\begin{equation}
\frac{d}{d\tau }\left( \frac{\partial L}{\partial \tilde{q}_{_{i}}}\right) -%
\frac{\partial L}{\partial q_{_{i}}}=0;\text{ }\,i=1,2,\cdots ,n.
\end{equation}%
would imply, with $m_{\circ }=1$, the $n$ EL-G equations of motion%
\begin{equation}
\frac{d}{d\tau }\tilde{q}_{_{i}}+\frac{\partial }{\partial q_{_{i}}}V(\vec{q}%
)=0,
\end{equation}%
in the generalized coordinates. Nevertheless, one should notice that for a
particle with $m_{\circ }\neq 1$ moving in a \emph{free} force field $V(\vec{%
q})=0\Longrightarrow d\tilde{q}_{_{j}}/d\tau =0$ the canonical momenta $%
m_{\circ }\tilde{q}_{_{j}}$ are conserved quantities ( in this particular
case) and serve as fundamental integrals (i.e., constants of motion).

At this point, we shall seek some sort of feasible invariance for PDM EL-I
(6) and PDM EL-II (7) with EL-G of (10). In so doing, we invest in the
one-dimensional nonlocal point transformation suggested by Mustafa \cite{38}
and extend/generalize it to fit into the current $n$-dimensional settings.
We, therefore, define%
\begin{equation}
\text{ \ }d\tau =f\left( \vec{x}\right) dt,\,dq_{_{i}}=\delta _{ij}\sqrt{%
g\left( \vec{x}\right) }\,dx_{_{j}}\Longrightarrow \frac{\partial q_{_{i}}}{%
\partial x_{_{j}}}=\delta _{ij}\sqrt{g\left( \vec{x}\right) },
\end{equation}%
where the functional structure of $f\left( \vec{x}\right) $ and $g\left( 
\vec{x}\right) $ shall be determined in the process below. This necessarily
means that the unit vectors in the direction of $q_{_{i}}$ are obtained as 
\begin{equation}
\hat{q}_{_{i}}=\frac{\sum\limits_{k=1}^{n}\left( \frac{\partial x_{_{k}}}{%
\partial q_{_{i}}}\right) \hat{x}_{k}}{\sqrt{\sum\limits_{k=1}^{n}\left( 
\frac{\partial x_{_{k}}}{\partial q_{_{i}}}\right) ^{2}}}=\frac{%
\sum\limits_{k=1}^{n}\delta _{ik}\hat{x}_{k}/\sqrt{g\left( \vec{x}\right) }}{%
\sqrt{\sum\limits_{k=1}^{n}\left( \delta _{ik}/\sqrt{g\left( \vec{x}\right) }%
\right) ^{2}}}\Longrightarrow \hat{q}_{_{i}}=\,\hat{x}_{_{i}};\text{ }%
i=1,2,\cdots ,n.
\end{equation}%
Under such settings, one obtains%
\begin{equation}
\text{\ }\tilde{q}_{_{j}}=\frac{\sqrt{g\left( \vec{x}\right) }}{f\left( \vec{%
x}\right) }\dot{x}_{_{j}}\,\Longrightarrow \frac{d}{d\tau }\text{\ }\tilde{q}%
_{_{j}}=\frac{\sqrt{g\left( \vec{x}\right) }}{f\left( \vec{x}\right) ^{2}}%
\left( \ddot{x}_{_{j}}+\dot{x}_{_{j}}\left[ \frac{\dot{g}\left( \vec{x}%
\right) }{2g\left( \vec{x}\right) }-\frac{\dot{f}\left( \vec{x}\right) }{%
f\left( \vec{x}\right) }\right] \right) .
\end{equation}%
and EL-G of (10) in turn implies the Euler-Lagrange equations%
\begin{equation}
\ddot{x}_{_{j}}+\left( \frac{\dot{g}\left( \vec{x}\right) }{2g\left( \vec{x}%
\right) }-\frac{\dot{f}\left( \vec{x}\right) }{f\left( \vec{x}\right) }%
\right) \dot{x}_{_{j}}+\left( \frac{f\left( \vec{x}\right) ^{2}}{g\left( 
\vec{x}\right) }\right) \,\partial _{x_{j}}V(\vec{q}\left( \vec{x}\right)
)=0;\,\,\,\,j=1,2,\cdots ,n.
\end{equation}%
This result is to be compared with PDM EL-I of (6) and PDM EL-II of (7).

The comparison between (14) and PDM EL-II of (7) is only possible for the
one-dimensional problems (i.e., for $n=1$). In this case, (7) collapses into
(6) for $i=1=n$. Nevertheless. for the multidimensional case $n\geq 2$, the
third term in (7) has no counterpart in (14). This would, in effect, make
the comparison incomplete/impossible and insecure. That is, for $n\geq 2$
the Euler-Lagrange equations (7) and (14) suggest that the invariance is
still far beyond reach under our PDM nonlocal point transformation. Hence,
we discard PDM EL-II of (7).

Whereas, the comparison between the second term of (14) and the second term
of the PDM EL-I of (6) immediately implies that the functional structure of $%
f\left( \vec{x}\right) $ and $g\left( \vec{x}\right) $ should necessarily be
the same as that of $m_{_{i}}\left( x_{i}\right) $ to allow exact
correspondence (i.e., $f\left( \vec{x}\right) =f\left( x_{i}\right) $ and $%
g\left( \vec{x}\right) =g\left( x_{i}\right) $) so that 
\begin{equation}
\frac{\dot{m}_{_{i}}\left( x_{i}\right) }{2m_{_{i}}\left( x_{i}\right) }=%
\frac{\dot{g}_{_{i}}\left( x_{i}\right) }{2g_{_{i}}\left( x_{i}\right) }-%
\frac{\dot{f}_{_{i}}\left( x_{i}\right) }{f_{_{i}}\left( x_{i}\right) }%
\Longleftrightarrow g_{_{i}}\left( x_{i}\right) =m_{_{i}}\left( x_{i}\right)
f_{_{i}}\left( x_{i}\right) ^{2}.
\end{equation}%
This would in turn make the third terms in (14) and (6) consistent (hence
the invariance is secured) to result in%
\begin{equation}
\frac{f_{_{i}}\left( x_{i}\right) ^{2}}{g_{_{i}}\left( x_{i}\right) }=\frac{1%
}{m_{_{i}}\left( x_{i}\right) }\Longleftrightarrow V\left( \vec{x}\right)
=V\left( \vec{q}\left( \vec{x}\right) \right) .
\end{equation}%
This would necessarily mean that the time $t$ is deformed/re-scaled in a
particular way for each coordinate $x_{i}$ so that our nonlocal point
transformation is summarized as%
\begin{equation}
\frac{d}{d\tau _{_{i}}}\left( \frac{\partial L}{\partial \tilde{q}_{_{i}}}%
\right) -\frac{\partial L}{\partial q_{_{i}}}=0\Longleftrightarrow \left\{ 
\begin{array}{c}
\partial q_{_{i}}/\partial x_{_{i}}=\sqrt{g_{_{i}}\left( x_{i}\right) }\,,%
\text{\ } \\ 
d\tau _{_{i}}\,/\medskip dt=f_{_{i}}\left( x_{i}\right)  \\ 
\,g_{_{i}}\left( x_{i}\right) =m_{_{i}}\left( x_{i}\right) f_{_{i}}\left(
x_{i}\right) ^{2}\medskip \medskip  \\ 
\text{ }V\left( \vec{x}\right) =V\left( \vec{q}\left( \vec{x}\right) \right)
\medskip  \\ 
\tilde{q}_{_{i}}\left( x_{i}\right) =\frac{1}{f_{_{i}}\left( x_{i}\right) }%
\dot{q}_{_{i}}\left( x_{i}\right) =\dot{x}_{_{i}}\sqrt{m_{_{i}}\left(
x_{i}\right) }%
\end{array}%
\right\} \Longleftrightarrow \frac{d}{dt}\left( \frac{\partial L_{I}}{%
\partial \dot{x}_{i}}\right) -\frac{\partial L_{I}}{\partial x_{i}}=0.
\end{equation}%
Where, the form of $f_{_{i}}\left( x_{i}\right) $ shall be determined by the
condition $V\left( \vec{x}\right) =V\left( \vec{q}\left( \vec{x}\right)
\right) $ for a given $m_{_{i}}\left( x_{i}\right) $. Consequently, one may
recast the PDM EL-I of (6) as%
\begin{equation}
\ddot{x}_{_{i}}+\left( \frac{m_{_{i}}^{\prime }\left( x_{i}\right) }{%
2m_{_{i}}\left( x_{i}\right) }\right) \,\dot{x}_{i}^{2}+\left( \frac{1}{%
m_{_{i}}\left( x_{i}\right) }\right) \,\partial _{x_{i}}V_{I}\left( \vec{x}%
\right) =0;\text{ }m_{_{i}}^{\prime }\left( x_{i}\right) =\frac{%
dm_{_{i}}\left( x_{i}\right) }{dx_{i}},
\end{equation}%
and consider it as the\emph{\ target} equation to be solved. We, therefore,
adopt PDM EL-I of (6) and the nonlocal point transformation (17) to proceed
with illustrative examples.

\section{Nonlinear $n$-dimensional PDM-oscillators}

Herein, we shall deal with nonlinear PDM-oscillators that are generated from
the force field 
\begin{equation}
V\left( \vec{q}\right) \medskip =\frac{1}{2}\sum\limits_{j=1}^{n}\omega
_{_{j}}^{2}q_{_{j}}^{2}
\end{equation}%
in the generalized coordinates and use the EL-G equations of (10) to yield
the $n$ EL-G equations of motion%
\begin{equation}
\frac{d}{d\tau _{_{i}}}\tilde{q}_{_{i}}+\frac{\partial }{\partial q_{_{i}}}V(%
\vec{q})=0\Longleftrightarrow \frac{d}{d\tau _{_{i}}}\tilde{q}_{_{i}}+\omega
_{_{i}}^{2}q_{_{i}}\,=0,
\end{equation}%
that admit exact solutions in the form of 
\begin{equation}
q_{_{i}}=B_{_{i}}\cos \left( \omega _{_{i}}\tau _{_{i}}+\varphi
_{_{i}}\right) .
\end{equation}%
We then choose our $q_{_{i}}\left( x_{i}\right) $ so that not only the
condition $V\left( \vec{x}\right) =V\left( \vec{q}\left( \vec{x}\right)
\right) $ is satisfied but also it is consistent with our nonlocal point
transformation (17) so that $\tilde{q}_{_{i}}\left( x_{i}\right) =\dot{x}%
_{_{i}}\sqrt{m_{_{i}}\left( x_{_{i}}\right) }$. The recipe is clear,
therefore.

Let us start with the potential force field,%
\begin{equation}
V_{I}\left( \vec{x}\right) =\frac{1}{2}\sum\limits_{i=1}^{n}m_{_{i}}\left(
x_{_{i}}\right) \,\omega _{_{i}}^{2}x_{_{i}}^{2}
\end{equation}%
in $L_{I}\left( \overrightarrow{x},\overrightarrow{\dot{x}};t\right) $ of
(3). This would in turn imply, with the condition $V\left( \vec{x}\right)
=V\left( \vec{q}\left( \vec{x}\right) \right) $, that 
\begin{equation}
q_{_{i}}\left( x_{_{i}}\right) =x_{_{i}}\sqrt{m_{_{i}}\left( x_{_{i}}\right) 
}.
\end{equation}%
In a straightforward manner one can immediately show that%
\begin{equation*}
q_{_{i}}\left( x_{_{i}}\right) =x_{_{i}}\sqrt{m_{_{i}}\left( x_{_{i}}\right) 
}\Longleftrightarrow \tilde{q}_{_{i}}\left( x_{_{i}}\right) =\frac{1}{%
f_{_{i}}\left( x_{_{i}}\right) }\dot{q}_{_{i}}\left( x_{_{i}}\right) =\dot{x}%
_{_{i}}\sqrt{m_{_{i}}\left( x_{_{i}}\right) },
\end{equation*}%
satisfies the nonlocal point transformation (17), where%
\begin{equation}
f_{_{i}}\left( x_{_{i}}\right) =1+\frac{m_{_{i}}^{\prime }\left(
x_{_{i}}\right) }{2m_{_{i}}\left( x_{_{i}}\right) }x_{_{i}}.
\end{equation}%
Consequently, the invariance between PDM EL-I of (6) and EL-G of (10) is
secured. Yet, one may choose $m_{_{i}}\left( x_{_{i}}\right) $ at will and
find the corresponding $f_{_{i}}\left( x_{_{i}}\right) $ in (24) and proceed
with the determination of the solutions for%
\begin{equation}
\ddot{x}_{_{i}}+\left( \frac{m_{_{i}}^{\prime }\left( x_{i}\right) }{%
2m_{_{i}}\left( x_{i}\right) }\right) \,\dot{x}_{i}^{2}+f_{_{i}}\left(
x_{_{i}}\right) \,\omega _{_{i}}^{2}x_{_{i}}=0.
\end{equation}%
This result is obtained either by the substitutions of (23) in (20) or by
the substitution of (22) in (18). We may now use different $m_{_{i}}\left(
x_{_{i}}\right) $ coordinates deformations and cast the following examples.

\subsection{Mathews-Lakshmanan type-I PDM-oscillators; coordinates
deformation $m_{_{i}}\left( x_{_{i}}\right) =\left( 1\pm \protect\lambda %
x_{_{i}}^{2}\right) ^{-1}$}

Let us consider a particle of rest $m_{\circ }$ moving in a an $n$%
-dimensional potential force field 
\begin{equation}
V_{I}\left( \vec{x}\right) =\frac{1}{2}m_{\circ }\sum\limits_{i=1}^{n}\frac{%
\,\omega _{_{i}}^{2}x_{_{i}}^{2}}{1\pm \lambda x_{_{i}}^{2}};\,\text{\ }%
m_{_{i}}\left( x_{_{i}}\right) =\frac{1}{1\pm \lambda x_{_{i}}^{2}},
\end{equation}%
and described by the $n$-dimensional \ Mathews-Lakshmanan type-I
PDM-oscillators Lagrangian%
\begin{equation}
L_{I}\left( \overrightarrow{x},\overrightarrow{\dot{x}};t\right) =\frac{1}{2}%
m_{\circ }\sum\limits_{i=1}^{n}\,\frac{\dot{x}_{_{j}}^{2}}{1\pm \lambda
x_{_{i}}^{2}}-\frac{1}{2}m_{\circ }\sum\limits_{i=1}^{n}\frac{\,\omega
_{_{i}}^{2}x_{_{i}}^{2}}{1\pm \lambda x_{_{i}}^{2}}.
\end{equation}%
This would necessarily imply that 
\begin{equation}
f_{_{i}}\left( x_{_{i}}\right) =1+\frac{m_{_{i}}^{\prime }\left(
x_{_{i}}\right) }{2m_{_{i}}\left( x_{_{i}}\right) }x_{_{i}}=\frac{1}{1\pm
\lambda x_{_{i}}^{2}}=m_{_{i}}\left( x_{_{i}}\right) .
\end{equation}%
Under such settings, obviously, the PDM EL-I of (25) would yield, with $%
m_{\circ }=1$, the $n$ Mathews-Lakshmanan type-I PDM-oscillators' equations
of motion%
\begin{equation}
\ddot{x}_{_{i}}\mp \left( \frac{\lambda x_{_{i}}}{1\pm \lambda x_{_{i}}^{2}}%
\right) \,\dot{x}_{i}^{2}+\left( \frac{1}{1\pm \lambda x_{_{i}}^{2}}\right)
\,\omega _{_{i}}^{2}x_{_{i}}=0;\text{ }i=1,2,\cdots ,n,
\end{equation}%
that admit exact solutions in the form of 
\begin{equation}
x_{_{i}}=A_{_{i}}\cos \left( \Omega _{_{i}}t+\varphi _{_{i}}\right) \,;\text{
\ }\Omega _{_{i}}^{2}=\frac{\omega _{_{i}}^{2}A_{_{i}}^{2}}{1\pm \lambda
A_{_{i}}^{2}}.
\end{equation}%
Then the total energy would eventually read%
\begin{equation}
E=\frac{1}{2}\sum\limits_{i=1}^{n}\text{\ }\Omega _{_{i}}^{2}A_{_{i}}^{2}=%
\frac{1}{2}\sum\limits_{i=1}^{n}\text{\ }\frac{\omega _{_{i}}^{2}A_{_{i}}^{2}%
}{1\pm \lambda A_{_{i}}^{2}}.
\end{equation}

\subsection{Nonlinear PDM-oscillators: Power-law type coordinates
deformation $m_{_{i}}\left( x_{_{i}}\right) =\protect\alpha ^{2}x_{_{i}}^{2%
\protect\upsilon }\,$}

Consider a particle of rest $m_{\circ }$ moving in a an $n$-dimensional
potential force field 
\begin{equation}
V_{I}\left( \vec{x}\right) =\frac{1}{2}m_{\circ }\sum\limits_{i=1}^{n}\left(
\alpha ^{2}x_{_{i}}^{2\upsilon }\right) \,\omega _{_{i}}^{2}x_{_{i}}^{2};\,%
\text{\ }m_{_{i}}\left( x_{_{i}}\right) =\alpha ^{2}x_{_{i}}^{2\upsilon },
\end{equation}%
where $\upsilon \neq -1$ (otherwise a dynamical collapse will occur, i.e., $%
q_{_{i}}\left( x_{_{i}}\right) =\alpha $ for $\upsilon =-1$), and hence
described by the $n$-dimensional PDM Lagrangian%
\begin{equation}
L_{I}\left( \overrightarrow{x},\overrightarrow{\dot{x}};t\right) =\frac{1}{2}%
m_{\circ }\sum\limits_{i=1}^{n}\left( \alpha ^{2}x_{_{i}}^{2\upsilon
}\right) \,\dot{x}_{_{j}}^{2}-\frac{1}{2}m_{\circ
}\sum\limits_{i=1}^{n}\left( \alpha ^{2}x_{_{i}}^{2\upsilon }\right)
\,\omega _{_{i}}^{2}x_{_{i}}^{2}.
\end{equation}%
In this case, the PDM EL-I of (25) would imply, with $m_{\circ }=1$ the $n$
nonlinear PDM-oscillators' equations of motion%
\begin{equation}
\ddot{x}_{_{i}}+\left( \frac{\upsilon }{x_{_{i}}}\right) \,\dot{x}%
_{i}^{2}+\left( 1+\upsilon \right) \,\omega _{_{i}}^{2}x_{_{i}}^{2}=0;\text{ 
}i=1,2,\cdots ,n,
\end{equation}%
that admit exact solutions in the form of 
\begin{equation}
x_{_{i}}=A_{_{i}}\left[ \cos \left( \Omega _{_{i}}t+\varphi _{_{i}}\right) %
\right] ^{1/\left( 1+\upsilon \right) }\,;\text{ \ }\Omega
_{_{i}}^{2}=\left( 1+\upsilon \right) ^{2}\omega _{_{i}}^{2},\,\upsilon \neq
-1.
\end{equation}%
The total energy would then read%
\begin{equation}
E=\frac{1}{2}\sum\limits_{i=1}^{n}\text{\ }\frac{\alpha ^{2}B_{_{i}}^{2}}{%
\left( 1+\upsilon \right) ^{2}}\,\Omega _{_{i}}^{2}=\frac{1}{2}%
\sum\limits_{i=1}^{n}\text{\ }\alpha ^{2}B_{_{i}}^{2}\,\omega _{_{i}}^{2};%
\text{ \ }B_{_{i}}=A_{_{i}}^{1/\left( 1+\upsilon \right) }.
\end{equation}

\subsection{Mathews-Lakshmanan type-II PDM-oscillators; coordinates
deformation $m_{_{i}}\left( x_{_{i}}\right) =\left( 1\pm \protect\lambda %
x_{_{i}}^{2}\right) ^{-1}$}

The $n$-dimensional Mathews-Lakshmanan PDM-oscillators are also possible to
be obtained by the consideration that the particle of rest $m_{\circ }$ is
moving in a an $n$-dimensional potential force field 
\begin{equation}
V_{I}\left( \vec{x}\right) =\frac{1}{2}m_{\circ }\sum\limits_{i=1}^{n}\frac{%
\,\omega _{_{i}}^{2}\eta _{_{i}}^{2}}{1\pm \lambda x_{_{i}}^{2}};\,\text{\ }%
m_{_{i}}\left( x_{_{i}}\right) =\frac{1}{1\pm \lambda x_{_{i}}^{2}},
\end{equation}%
where $\eta _{_{i}}$'s are constants and hence such a particle is described
by the $n$-dimensional \ Mathews-Lakshmanan type-II PDM-oscillators'
Lagrangian%
\begin{equation}
L_{I}\left( \overrightarrow{x},\overrightarrow{\dot{x}};t\right) =\frac{1}{2}%
m_{\circ }\sum\limits_{i=1}^{n}\,\frac{\dot{x}_{_{j}}^{2}}{1\pm \lambda
x_{_{i}}^{2}}-\frac{1}{2}m_{\circ }\sum\limits_{i=1}^{n}\frac{\,\omega
_{_{i}}^{2}\eta _{_{i}}^{2}}{1\pm \lambda x_{_{i}}^{2}}.
\end{equation}%
Under such settings, equation (17) suggests that 
\begin{equation}
q_{_{i}}\left( x_{_{i}}\right) =\eta _{_{i}}\sqrt{m_{_{i}}\left(
x_{_{i}}\right) }\Longleftrightarrow \dot{q}_{_{i}}\left( x_{_{i}}\right)
=\left( \eta _{_{i}}\frac{m_{_{i}}^{\prime }\left( x_{_{i}}\right) }{%
2m_{_{i}}\left( x_{_{i}}\right) }\right) \sqrt{m_{_{i}}\left(
x_{_{i}}\right) }\dot{x}_{_{i}}\Longleftrightarrow f_{_{i}}\left(
x_{_{i}}\right) =\eta _{_{i}}\frac{m_{_{i}}^{\prime }\left( x_{_{i}}\right) 
}{2m_{_{i}}\left( x_{_{i}}\right) }.
\end{equation}%
Consequently, the EL-G of (20) would, with $m_{\circ }=1$, imply that%
\begin{equation}
\ddot{x}_{_{i}}+\left( \frac{m_{_{i}}^{\prime }\left( x_{i}\right) }{%
2m_{_{i}}\left( x_{i}\right) }\right) \,\dot{x}_{i}^{2}+f_{_{i}}\left(
x_{_{i}}\right) \eta _{_{i}}\,\omega _{_{i}}^{2}=0.
\end{equation}%
Obviously, this result would yield the $n$ dynamical Mathews-Lakshmanan
type-II equations of motion%
\begin{equation}
\ddot{x}_{_{i}}\mp \left( \frac{\lambda x_{_{i}}}{1\pm \lambda x_{_{i}}^{2}}%
\right) \,\dot{x}_{i}^{2}+\left( \frac{\mp \lambda \eta _{_{i}}^{2}}{1\pm
\lambda x_{_{i}}^{2}}\right) \,\omega _{_{i}}^{2}x_{_{i}}=0;\text{ }%
i=1,2,\cdots ,n.
\end{equation}%
It is clear that for $\lambda =\mp 1/\eta _{_{i}}^{2}$ these equations of
motion collapse into those obtained for Mathews-Lakshmanan type-I
oscillators in (29) and, therefore, inherit the corresponding solutions in
(30) and (31).

\subsection{PDM-Morse oscillators; coordinates deformation $m_{_{i}}\left(
x_{_{i}}\right) =\exp (2\protect\zeta x_{_{i}})$}

We now consider the particle of rest $m_{\circ }$ moving in an $n$%
-dimensional deformed Morse-like oscillator force field%
\begin{equation}
V_{I}\left( \vec{x}\right) =\frac{1}{2}m_{\circ
}\sum\limits_{i=1}^{n}m_{_{i}}\left( x_{_{i}}\right) \,\omega _{_{i}}^{2}%
\left[ 1-\exp \left( -\zeta _{_{i}}x_{_{i}}\right) \right] ^{2}.
\end{equation}%
The Lagrangian describing such a system is given by%
\begin{equation}
L_{I}\left( \overrightarrow{x},\overrightarrow{\dot{x}};t\right) =\frac{1}{2}%
m_{\circ }\sum\limits_{i=1}^{n}\,m_{_{i}}\left( x_{_{i}}\right) \dot{x}%
_{_{i}}^{2}-\frac{1}{2}m_{\circ }\sum\limits_{i=1}^{n}m_{_{i}}\left(
x_{_{i}}\right) \,\omega _{_{i}}^{2}\left[ 1-\exp \left( -\zeta
_{_{i}}x_{_{i}}\right) \right] ^{2}.
\end{equation}%
In this case, with $m_{\circ }=1$,\ equation (16) mandates the assumption
that%
\begin{equation}
q_{_{i}}\left( x_{_{i}}\right) =\sqrt{m_{_{i}}\left( x_{_{i}}\right) }\left[
1-\exp \left( -\zeta _{_{i}}x_{_{i}}\right) \right] ,
\end{equation}%
to obtain%
\begin{equation}
\dot{q}_{_{i}}\left( x_{_{i}}\right) =\sqrt{m_{_{i}}\left( x_{_{i}}\right) }%
\dot{x}_{_{i}}\left[ \frac{m_{_{i}}^{\prime }\left( x_{i}\right) }{%
2m_{_{i}}\left( x_{i}\right) }+\left( \zeta _{_{i}}-\frac{m_{_{i}}^{\prime
}\left( x_{i}\right) }{2m_{_{i}}\left( x_{i}\right) }\right) \exp \left(
-\zeta _{_{i}}x_{_{i}}\right) \right] .
\end{equation}%
Nevertheless, our nonlocal point transformation in (17) yields%
\begin{equation}
\dot{q}_{_{i}}\left( x_{_{i}}\right) =_{_{i}}f_{_{i}}\left( x_{_{i}}\right) 
\sqrt{m_{_{i}}\left( x_{_{i}}\right) }\dot{x}\Longleftrightarrow
f_{_{i}}\left( x_{_{i}}\right) =\left[ \frac{m_{_{i}}^{\prime }\left(
x_{i}\right) }{2m_{_{i}}\left( x_{i}\right) }+\left( \zeta _{_{i}}-\frac{%
m_{_{i}}^{\prime }\left( x_{i}\right) }{2m_{_{i}}\left( x_{i}\right) }%
\right) \exp \left( -\zeta _{_{i}}x_{_{i}}\right) \right] ,
\end{equation}%
which, for $m_{_{i}}\left( x_{_{i}}\right) =\exp \left( 2\zeta
_{_{i}}x_{_{i}}\right) $ implies the simplistic forms%
\begin{equation}
\dot{q}_{_{i}}\left( x_{_{i}}\right) =\zeta _{_{i}}\sqrt{m_{_{i}}\left(
x_{_{i}}\right) }\dot{x}\Longleftrightarrow _{_{i}}f_{_{i}}\left(
x_{_{i}}\right) =\zeta _{_{i}}.
\end{equation}%
At this point one should notice that the choice of $m_{_{i}}\left(
x_{_{i}}\right) =\exp \left( 2\zeta _{_{i}}x_{_{i}}\right) $ is not unique
but rather keeps the problem simple and straightforward. Under such
settings, equation (20) would result%
\begin{equation}
\ddot{x}_{_{i}}+\zeta _{_{i}}\,\dot{x}_{i}^{2}+\,\omega _{_{i}}^{2}\zeta
_{_{i}}\left( 1-\exp \left( -\zeta _{_{i}}x_{_{i}}\right) \right) =0.
\end{equation}%
Equation (21) also suggests, along with (44), that 
\begin{equation}
q_{_{i}}\left( x_{_{i}}\right) =B_{_{i}}\cos \left( \omega _{_{i}}\tau
_{i}+\varphi _{_{i}}\right) \Longleftrightarrow q_{_{i}}\left(
x_{_{i}}\right) =\exp \left( \zeta _{_{i}}x_{_{i}}\right) -1.
\end{equation}%
This would, in turn, imply that%
\begin{equation}
x_{_{i}}=\frac{1}{\zeta _{_{i}}}\ln \left[ 1+B_{_{i}}\cos \left( \zeta
_{_{i}}\omega _{_{i}}t+\varphi _{_{i}}\right) \right] 
\end{equation}%
as the exact solutions for PDM EL-I of (48). The total energy, moreover,
reads%
\begin{equation}
E=\frac{1}{2}\sum\limits_{i=1}^{n}\,\omega _{_{i}}^{2}B_{_{i}}^{2}.
\end{equation}

\section{Nonlinear isotonic $n$-dimensional PDM-oscillators}

A particle of rest $m_{\circ }$ moving in a nonlinear isotonic $n$%
-dimensional oscillator force field%
\begin{equation}
V\left( \vec{q}\right) \medskip =\frac{1}{2}m_{\circ
}\sum\limits_{j=1}^{n}\left( \omega _{_{j}}^{2}q_{_{j}}^{2}+\frac{\kappa
_{_{j}}}{q_{_{j}}^{2}}\right) ,
\end{equation}%
is described by the so called Smorodinsky-Winternitz type (c.f., e.g., \cite%
{29,38,39,40,41,42,43}) $n$-dimensional Lagrangian%
\begin{equation}
L\left( \overrightarrow{q},\overrightarrow{\tilde{q}};\tau \right) =\frac{1}{%
2}m_{\circ }\sum\limits_{j=1}^{n}\tilde{q}_{_{j}}^{2}-\frac{1}{2}m_{\circ
}\sum\limits_{j=1}^{n}\left( \omega _{_{j}}^{2}q_{_{j}}^{2}+\frac{\kappa
_{_{j}}}{q_{_{j}}^{2}}\right) .
\end{equation}%
This would imply, with  $m_{\circ }=1$, the $n$ EL-G equations\ of motion%
\begin{equation}
\frac{d}{d\tau _{_{i}}}\tilde{q}_{_{i}}+\frac{\partial }{\partial q_{_{i}}}V(%
\vec{q})=0\Longleftrightarrow \frac{d}{d\tau _{_{i}}}\tilde{q}_{_{i}}+\omega
_{_{i}}^{2}q_{_{i}}-\frac{\kappa _{_{i}}}{q_{_{i}}^{3}}=0,
\end{equation}%
that admit exact solutions in the form of%
\begin{equation}
q_{_{i}}=\frac{1}{\omega _{_{i}}C_{_{i}}}\sqrt{\omega
_{_{i}}^{2}C_{_{i}}^{4}\sin ^{2}\left( \omega _{_{i}}\tau _{_{i}}+\sigma
_{_{i}}\right) +\kappa _{_{i}}\cos ^{2}\left( \omega _{_{i}}\tau
_{_{i}}+\sigma _{_{i}}\right) }
\end{equation}

Next, let us consider the potential force field%
\begin{equation}
V_{I}\left( \vec{x}\right) =\frac{1}{2}\sum\limits_{i=1}^{n}\left(
m_{_{i}}\left( x_{_{i}}\right) \,\omega _{_{i}}^{2}x_{_{i}}^{2}+\frac{\kappa
_{_{i}}}{m_{_{i}}\left( x_{_{i}}\right) x_{i}^{2}}\right) 
\end{equation}%
in $L_{I}\left( \overrightarrow{x},\overrightarrow{\dot{x}};t\right) $ of
(3). This would in turn imply, with the condition $V\left( \vec{x}\right)
=V\left( \vec{q}\left( \vec{x}\right) \right) $, that 
\begin{equation}
q_{_{i}}\left( x_{_{i}}\right) =x_{_{i}}\sqrt{m_{_{i}}\left( x_{_{i}}\right) 
}\Longleftrightarrow \tilde{q}_{_{i}}\left( x_{_{i}}\right) =\frac{1}{%
f_{_{i}}\left( x_{_{i}}\right) }\dot{q}_{_{i}}\left( x_{_{i}}\right) =\dot{x}%
_{_{i}}\sqrt{m_{_{i}}\left( x_{_{i}}\right) }\Longleftrightarrow
f_{_{i}}\left( x_{_{i}}\right) =1+\frac{m_{_{i}}^{\prime }\left(
x_{_{i}}\right) }{2m_{_{i}}\left( x_{_{i}}\right) }x_{_{i}}.
\end{equation}%
In a straightforward manner, one can immediately show that the PDM EL-I of
(18) or EL-G (54) would yield the $n$ dynamical PDM Euler-Lagrange equations
of motion 
\begin{equation}
\ddot{x}_{_{i}}+\left( \frac{m_{_{i}}^{\prime }\left( x_{i}\right) }{%
2m_{_{i}}\left( x_{i}\right) }\right) \,\dot{x}_{i}^{2}+f_{_{i}}\left(
x_{_{i}}\right) \left( \,\omega _{_{i}}^{2}x_{_{i}}-\frac{\kappa _{_{i}}}{%
m_{_{i}}\left( x_{_{i}}\right) ^{2}x_{i}^{3}}\right) =0.
\end{equation}%
Such nonlinear isotonic $n$-dimensional PDM-oscillators equations of motion
(54) and (58) are to be used to reflect on the exact solutions of the
following examples.

\subsection{Smorodinsky-Winternitz type-I PDM-oscillators; coordinates
deformation $m_{_{i}}\left( x_{_{i}}\right) =\left( 1\pm \protect\lambda %
x_{_{i}}^{2}\right) ^{-1}$}

Let us consider a particle of rest $m_{\circ }$ moving in a an $n$%
-dimensional potential force field 
\begin{equation}
V_{I}\left( \vec{x}\right) =\frac{1}{2}m_{\circ }\sum\limits_{i=1}^{n}\left( 
\frac{\,\omega _{_{i}}^{2}x_{_{i}}^{2}}{1\pm \lambda x_{_{i}}^{2}}+\frac{%
\left( 1\pm \lambda x_{_{i}}^{2}\right) \kappa _{_{i}}}{x_{_{i}}^{2}}\right)
;\,\text{\ }m_{_{i}}\left( x_{_{i}}\right) =\frac{1}{1\pm \lambda
x_{_{i}}^{2}},
\end{equation}%
and hence is described by the $n$-dimensional \ Smorodinsky-Winternitz
type-I PDM-oscillators' Lagrangian%
\begin{equation}
L_{I}\left( \overrightarrow{x},\overrightarrow{\dot{x}};t\right) =\frac{1}{2}%
m_{\circ }\sum\limits_{i=1}^{n}\,\frac{\dot{x}_{_{i}}^{2}}{1\pm \lambda
x_{_{i}}^{2}}-\frac{1}{2}m_{\circ }\sum\limits_{i=1}^{n}\left( \frac{%
\,\omega _{_{i}}^{2}x_{_{i}}^{2}}{1\pm \lambda x_{_{i}}^{2}}+\frac{\kappa
_{_{i}}\,\left( 1\pm \lambda x_{_{i}}^{2}\right) }{x_{_{i}}^{2}}\right) .
\end{equation}%
This would necessarily mean that 
\begin{equation}
f_{_{i}}\left( x_{_{i}}\right) =m_{_{i}}\left( x_{_{i}}\right) =\frac{1}{%
1\pm \lambda x_{_{i}}^{2}}.
\end{equation}%
Under such settings, obviously, the PDM EL-I of (58) would yield, with $%
m_{\circ }=1$, the $n$ dynamical Smorodinsky-Winternitz type-I
PDM-oscillators' equations of motion%
\begin{equation}
\ddot{x}_{_{i}}\mp \frac{\lambda x_{_{i}}\dot{x}_{i}^{2}}{1\pm \lambda
x_{_{i}}^{2}}\,+\frac{\omega _{_{i}}^{2}x_{_{i}}}{1\pm \lambda x_{_{i}}^{2}}%
\,-\frac{\kappa _{_{i}}}{x_{i}^{3}}\left( 1\pm \lambda x_{_{i}}^{2}\right)
=0.
\end{equation}%
The exact solution of which is%
\begin{equation}
x_{_{i}}=\frac{1}{\Omega _{_{i}}C_{_{i}}}\sqrt{\Omega
_{_{i}}^{2}C_{_{i}}^{4}\sin ^{2}\left( \Omega _{_{i}}t+\sigma _{_{i}}\right)
+\kappa _{_{i}}\cos ^{2}\left( \Omega _{_{i}}t+\sigma _{_{i}}\right) }%
;\,\,\omega _{_{i}}^{2}=\left( 1\pm \lambda C_{_{i}}^{2}\right) \left[
\Omega _{_{i}}^{2}\pm \frac{\lambda \kappa _{_{i}}}{C_{_{i}}^{2}}\right] 
\end{equation}

\subsection{Smorodinsky-Winternitz type-II PDM-oscillators; Power-law type
coordinates deformation}

We now consider a power-law type coordinate deformation%
\begin{equation}
m_{_{i}}\left( x_{_{i}}\right) =\beta ^{2}x_{_{i}}^{2\left( \eta -1\right)
}\,;\eta \neq 1
\end{equation}%
in the potential force field of (56), where $\eta \neq 1$ (otherwise for $%
\eta =1$ we retrieve the readily known constant mass settings) to obtain $n$
PDM Euler-Lagrange equations of motion \ 
\begin{equation}
\ddot{x}_{_{i}}+\left( \frac{\eta -1}{x_{_{i}}}\right) \,\dot{x}%
_{_{i}}^{2}+\eta \,\left( \,\omega _{_{i}}^{2}x_{_{i}}-\frac{\kappa _{_{i}}}{%
\beta ^{4}x_{_{i}}^{4\left( \eta -1\right) }x_{_{i}}^{3}}\right) =0.
\end{equation}%
In this case, $f_{_{i}}\left( x_{_{i}}\right) =\eta ,$ and the solutions of
the PDM Euler-Lagrange equations (65) read%
\begin{equation}
x_{_{i}}=\left[ \frac{1}{\beta \eta \,\omega _{_{i}}C_{_{i}}}\sqrt{\eta
^{2}\omega _{_{i}}^{2}C_{_{i}}^{4}\sin ^{2}\left( \,\omega _{_{i}}\eta
t+\sigma _{_{i}}\right) +\kappa _{_{i}}\cos ^{2}\left( \,\omega _{_{i}}\eta
t+\sigma _{_{i}}\right) }\right] ^{1/\eta }.
\end{equation}%
Which satisfies (65) if and only if%
\begin{equation}
\eta ^{2}-1=0\Longleftrightarrow \eta =-1\,;\,\eta \neq 1.
\end{equation}%
otherwise trivial solutions are obtained.

\section{Concluding Remarks}

In the current methodical proposal, we have introduced an $n$-dimensional
extension of the very recent one-dimensional nonlocal point transformation
recipe for PDM-Lagrangians by Mustafa \cite{38}. We have used two
PDM-Lagrangian models $L_{I}\left( \overrightarrow{x},\overrightarrow{\dot{x}%
};t\right) $ and $L_{II}\left( \overrightarrow{x},\overrightarrow{\dot{x}}%
;t\right) $ of (3) \ and (4), respectively, along with a standard textbook
Lagrangian (i.e., for constant mass) $L\left( \overrightarrow{q},%
\overrightarrow{\tilde{q}};\tau \right) $ of (8) in the generalized
coordinates. Using some $n$-dimensional nonlocal point transformation
(11)-(16), we have tested/experimented the invariance feasibility of the
corresponding PDM Euler-Lagrange equations (i.e., PDM EL-I of (6) for $%
L_{I}\left( \overrightarrow{x},\overrightarrow{\dot{x}};t\right) $ and PDM
EL-II of (7) for $L_{II}\left( \overrightarrow{x},\overrightarrow{\dot{x}}%
;t\right) $) with the standard Euler-Lagrange equation EL-G of (10) for $%
L\left( \overrightarrow{q},\overrightarrow{\tilde{q}};\tau \right) $. We
have observed that, while the invariance of PDM EL-I of (6) is secured and
is shown feasible, the invariance of PDM EL-II of (7) is obtained impossible
for $n\geq 2$ (i.e., for more than one--dimensional case). We, thereinafter,
adopted the PDM-Lagrangian model $L_{I}\left( \overrightarrow{x},%
\overrightarrow{\dot{x}};t\right) $ and its corresponding PDM EL-I of (6)
along with our $n$-dimensional nonlocal point transformation recipe,
summarized in (17), in order to insure the textbook invariance in the
process. This kind of Euler-Lagrange invariance, nevertheless, facilitates
exact solutions for problems with coordinates deformations that renders the
mass position-dependent (so to speak) as documented in the above
illustrative examples.

Yet, the $n$-dimensional PDM Euler-Lagrange equations that may look
complicated and/or insoluble problems turnout to admit simple and
straightforward solutions within the current nonlocal point transformation
proposal settings (17). In the nonlinear $n$-dimensional PDM-oscillators,
for example, the Mathews-Lakshmanan type-I oscillators of (29), admit exact
solutions that are extracted through the substitution 
\begin{equation}
q_{_{i}}=x_{_{i}}\sqrt{m_{_{i}}\left( x_{_{i}}\right) }=\frac{x_{_{i}}}{%
\sqrt{1\pm \lambda x_{_{i}}^{2}}}\Longleftrightarrow q_{_{i}}^{2}=\frac{%
x_{_{i}}^{2}}{1\pm \lambda x_{_{i}}^{2}}\Longleftrightarrow x_{_{i}}^{2}=%
\frac{q_{_{i}}^{2}}{1\mp \lambda q_{_{i}}^{2}}.
\end{equation}%
Which suggests that $x_{_{i}}$ and $q_{_{i}}$ have similar functional
structure and implies%
\begin{equation}
q_{_{j}}=A_{_{j}}\cos \left( \omega _{_{j}}\tau _{_{j}}+\varphi
_{_{i}}\right) \Longleftrightarrow x_{_{i}}=A_{_{j}}\cos \left( \Omega
_{_{j}}t+\varphi _{_{i}}\right) ;\text{ \ }\Omega _{_{i}}^{2}=\frac{\omega
_{_{i}}^{2}A_{_{i}}^{2}}{1\pm \lambda A_{_{i}}^{2}},
\end{equation}%
where $\Omega _{_{j}}$ is determined so that the dynamical equations in (29)
are satisfied. Whereas, for the power-law deformation of (34), the exact
solutions are extracted through the relations 
\begin{equation}
q_{_{i}}=x_{_{i}}\sqrt{m_{_{i}}\left( x_{_{i}}\right) }=\alpha
x_{_{i}}^{\alpha +1}\Longleftrightarrow x_{_{i}}=\left[ \frac{q_{_{i}}}{%
\alpha }\right] ^{1/\left( \alpha +1\right) }\Longleftrightarrow
x_{_{i}}=A_{_{i}}\left[ \cos \left( \Omega _{_{i}}t+\varphi _{_{i}}\right) %
\right] ^{1/\left( 1+\upsilon \right) };\text{ \ }\Omega _{_{i}}^{2}=\left(
1+\upsilon \right) ^{2}\omega _{_{i}}^{2}
\end{equation}%
and so on so forth. Similar arguments are also used to extract exact
solutions for the nonlinear isotonic $n$-dimensional PDM-oscillators
(documented in the Smorodinsky-Winternitz type-I oscillators of (62) and
Smorodinsky-Winternitz type-II oscillators of (65) illustrative examples).

Finally, as long as one is dealing with Lagrangians and/or Hamiltonians in
more than one dimension, the notion about their \emph{superintegrability}
(c.f., e.g., \cite{39,40,41,42,43} and related references cited therein) is
unavoidable in the process. Although \emph{superintegrability} lies far
beyond the scope of our current methodical proposal, there is no harm in
recollecting that a Lagrangian (likewise, a Hamiltonian) system is said to
be \emph{superintegrable} if it admits the Liouville-Arnold sense of
integrability and introduces more constants of motion (also called integrals
of motion) than the degrees of freedom the system is moving within. \
Notably,\ our \emph{reference} Lagrangians in the generalized coordinates
are within the general form 
\begin{equation}
L\left( \overrightarrow{q},\overrightarrow{\tilde{q}};\tau \right) =\frac{1}{%
2}m_{\circ }\sum\limits_{j=1}^{n}\tilde{q}_{_{j}}^{2}-\left( \frac{1}{2}%
\sum\limits_{j=1}^{n}\omega _{j}^{2}q_{_{j}}^{2}+\frac{1}{2}%
\sum\limits_{j=1}^{n}\frac{k_{_{j}}}{q_{_{j}}^{2}}\right) ;\text{ \ }\tilde{q%
}_{_{j}}=\frac{dq_{_{j}}}{d\tau };\text{ }\,j=1,\cdots ,n.
\end{equation}%
Which clearly suggests that they are \emph{superintegrable}. Consequently,
our \emph{target} PDM-Lagrangians%
\begin{equation}
L_{I}\left( \overrightarrow{x},\overrightarrow{\dot{x}};t\right) =\frac{1}{2}%
m_{\circ }\sum\limits_{j=1}^{n}m_{_{j}}\left( x_{_{j}}\right) \,\dot{x}%
_{_{j}}^{2}-V_{I}\left( \vec{x}\right) ;\text{ }\,\,\,j=1,2,\cdots ,n\in 
\mathbb{N}
,
\end{equation}%
are at least\emph{\ pseudo-superintegrable} for they are mapped (through our
nonlocal point transformation) into \emph{superintegrable} Lagrangians in
the generalized coordinates. Yet, the \emph{superintegrability} of such
PDM-Lagrangians can also be studied. 


\newpage 

\begin{thebibliography}{99}
\bibitem{1} O. von Roos, Phys. Rev. \textbf{B 27 }(1983) 7547.

\bibitem{2} A. de Souza Dutra, C A S Almeida, Phys Lett. \textbf{A 275}
(2000) 25.

\bibitem{3} O. Mustafa, J Phys \textbf{A}: Math. Theor.\textbf{52 (}2019%
\textbf{) }148001.

\bibitem{4} O. Mustafa, S. H. Mazharimousavi, Phys. Lett. \textbf{A 358}
(2006) 259.

\bibitem{5} O. Mustafa, Z. Algadhi, Eur. Phys. J Plus (2019) in press, 
\textbf{"}Position-dependent mass momentum operator and minimal coupling:
point canonical transformation and isospectrality" (arXiv:1806.02983).

\bibitem{6} B. Bagchi, A. Banerjee, C. Quesne, V. M. Tkachuk, J. \ Phys. 
\textbf{A}: Math. Gen. \textbf{38} (2005) 2929.

\bibitem{7} O. Mustafa, S. H. Mazharimousavi, Phys. Lett. \textbf{A 357}
(2006) 295.

\bibitem{8} O. Mustafa, J Phys \textbf{A}: Math. Theor. \textbf{44 (}2011%
\textbf{) }355303.

\bibitem{9} B. Bagchi, P. Gorain, C. Quesne and R. Roychoudhury, Mod. Phys.
Lett. \textbf{A 19} (2004) 2765.

\bibitem{10} O. Mustafa, S. H. Mazharimousavi, Int. J. Theor. Phys \ \textbf{%
46} (2007) 1786.

\bibitem{11} S. Cruz y Cruz, O Rosas-Ortiz, J Phys \textbf{A}: Math. Theor. 
\textbf{42} (2009) 185205.

\bibitem{12} S. Cruz y Cruz, J. Negro and L.M. Nieto, Phys. Lett. \textbf{A} 
\textbf{369,} (2007) 400.

\bibitem{13} S. Cruz y Cruz, J. Negro and L.M. Nieto, Journal of Physics:
Conference Series \textbf{128, }(2008)\textbf{\ }012053.

\bibitem{14} S. Cruz y Cruz, O Rosas-Ortiz, SIGMA \textbf{9, }(2013)\textbf{%
\ }004.

\bibitem{15} S. Ghosh and S. K. Modak, Phys. Lett. A \textbf{373,} (2009)
1212.

\bibitem{16} B. Bagchi, S. Das, S. Ghosh and S. Poria, J. Phys. \textbf{A}:
Math. Theor. \textbf{46,} (2013) 032001.

\bibitem{17} S. H. Mazharimousevi, O. Mustafa, Phys. Scr. \textbf{87} (2013)
055008.

O. Mustafa; arXiv:1208.2109:\textbf{\ }Comment on the "Classical and quantum
position-dependent mass harmonic oscillators" and ordering-ambiguity
resolution.

\bibitem{18} P. M. Mathews, M. Lakshmanan, Quart. Appl. Math. \textbf{32 }%
(1974)\textbf{\ }215.

\bibitem{19} A. Venkatesan, M. Lakshmanan, Phys. Rev. \textbf{E} \textbf{55}
(1997) 5134.

\bibitem{20} J. F. Cari\~{n}ena, M. F. Ra\~{n}ada, M. Santander, Regul.
Chaotic Dyn. \textbf{10} (2005) 423.

\bibitem{21} A. Bhuvaneswari, V. K. Chandrasekar, M. Santhilvelan, M.
Lakshmanan, J. Math. Phys. \textbf{53} (2012) 073504.

\bibitem{22} O. Mustafa, J. Phys. A: Math. Theor.\textbf{\ 46 }(2013) 368001.

\bibitem{23} A. K. Tiwari, S. N. Pandey, M. Santhilvelan, M. Lakshmanan, J.
Math. Phys. \textbf{54} (2013) 053506.

\bibitem{24} M. Lakshmanan, V. K. Chandrasekar, Eur. Phys J. ST \textbf{222}
(2013) 665.

\bibitem{25} Z. E. Musielak, J. Phys. \textbf{A}: Math. Theor. \textbf{41,}
(2008) 055205.

\bibitem{26} C. Quesne, J. Math. Phys. \textbf{56} (2015) 012903.

\bibitem{27} C. Quesne, V. M. Tkachuk, J. Phys. \textbf{A 37} (2004) 4267.

\bibitem{28} B. Bagchi, A. Banerjee, C. Quesne, V. M. Tkachuk, J. Phys. 
\textbf{A 38} (2005) 2929.

\bibitem{29} J. F. Cari\~{n}ena, M. F. Ra\~{n}ada, M. Santander, SIGMA 
\textbf{3} (2007) 030.

\bibitem{30} J. F. Cari\~{n}ena, M. F. Ra\~{n}ada, M. Santander, Ann. Phys. 
\textbf{322 }(2007) 434.

\bibitem{31} J. F. Cari\~{n}ena, M. F. Ra\~{n}ada, M. Santander, Rep. Math.
Phys. \textbf{54} (2004) 285.

\bibitem{32} C. Muriel, J. L. Romero, J. Phys. \textbf{A}: Math. Theor. 
\textbf{43} (2010) 434025.

\bibitem{33} R. G. Pradeep, V. K. Chandrasekar, M. Santhilvelan, M.
Lakshmanan, J. Math. Phys. \textbf{50} (2009) 052901.

\bibitem{34} N. Euler, M. Euler, J. Nonlinear Math. Phys. \textbf{11} (2004)
399.

\bibitem{35} W. -H. Steeb, "\textit{Invertible Point Transformation and
Nonlinear Differential Equations", }World Scientific, Singapore, 1993.

\bibitem{36} K. S. Govinder, P. G. L. Leach, J. Math. Anal. Appl. \textbf{%
287 }(2003) 399.

\bibitem{37} N. Euler, T. Wolf, P. G. L. Leach, M. Euler, Acta Appl. Math. 
\textbf{76} (2003) 89.

\bibitem{38} O. Mustafa, J. Phys. \textbf{A}; Math. Theor. \textbf{48}
(2015) 225206.

\bibitem{39} M. Ranada, M. A. Rodrigues, and M Santander, J. Math. Phys. 
\textbf{51}, 042901 (2010).

\bibitem{40} M. A. Rodrigues, p. Tempesta, and P. Winternitz, Phys. Rev. E 
\textbf{78} 046608 (2008).

\bibitem{41} M. Ranada, J. Math. Phys. \textbf{57}, 052703 (2016).

\bibitem{42} F. Tremblay, A. Turbiner, P. Winternitz, J. Phys. \textbf{A};
Math. Theor. \textbf{43} (2010) 015202; J. Phys. A \textbf{43}: Math. Theor.
(2010) 015202,

\bibitem{43} J. F. Cari\~{n}ena, F. J. Herranz, M. F. Ra\~{n}ada, J. Math.
Phys. \textbf{58} (2017) 022701.
\end{thebibliography}
\end{document}